\documentclass[12pt]{iopart}

\usepackage{iopams} 
\usepackage{graphicx}% Include figure files
\begin{document}

\title{STAR's measurement of Long-range forward-backward multiplicity
            correlations as the signature of ``dense partonic matter'' in
            the Heavy Ion collisions at $\sqrt{s_{NN}}=$200 GeV.}

\author{Brijesh K Srivastava( for the STAR Collaboration)$^{\rm a}$}

\address{$^{\rm a}$Department of Physics , Purdue University,
West Lafayette, Indiana-47907, USA}
\ead{brijesh@physics.purdue.edu}

\begin{abstract}

 Forward-backward multiplicity correlations have been measured with the
STAR detector for Au+Au, Cu+Cu and {\it p+p} collisions at $\sqrt{s_{NN}}$ = 200 GeV. A strong, long-range
correlation is observed for central heavy ion collisions that vanishes
in semi-peripheral events and {\it pp} collisions. There is no apparent
scaling of correlation strength with the number of participants involved in the collision. Both the
Dual Parton Model and the Color Glass condensate indicate that the
long range correlations are due to multiple parton interactions. This suggests
that the dense partonic matter might have been created in mid-central and central Au+Au
collisions at $\sqrt{s_{NN}}$ = 200 GeV.
\end{abstract}

%Uncomment for PACS numbers title message
%\pacs{00.00, 20.00, 42.10}
% Keywords required only for MST, PB, PMB, PM, JOA, JOB? 
%\vspace{2pc}
%\noindent{\it Keywords}: Article preparation, IOP journals
% Uncomment for Submitted to journal title message
%\submitto{\JPA}
% Comment out if separate title page not required
%\maketitle
The investigation of high energy nucleus-nucleus collisions provides a unique
tool to study the properties of hot and dense matter. The motivation is drawn
from lattice QCD calculations, which predicts a phase transition from
hadronic matter to a system of deconfined quarks and gluons (QGP) at high
temperature \cite{karsch}. The study of event-by-event correlations and fluctuations provides  a probe to explore such transition in the search for
the QGP. In particular the measurement of particle correlations has been suggested as a method to search for the existence of a phase transition in
ultra-relativistic heavy ion collisions \cite{rajgopal}.

It has been suggested that long-range rapidity correlations (LRC) might be enhanced
in hadron-nucleus and nucleus-nucleus interactions, compared to hadron-hadron scattering at the same energy \cite{capela1,capela2}. The presence  of
long range correlations implies the
existence of multiple inelastic collisions and provides a test of the multiple scattering models\cite{capela2}.
The Color Glass Condensate also predicts the large scale rapidity correlations in heavy ion collisions \cite{larry}.

The correlation strength is defined by the dependence of the average charged particle multiplicity in the backward hemisphere $\langle N_{b}\rangle$, on the
event multiplicity in the forward hemisphere $N_{f}$, $\langle N_{b}\rangle$=a+$b$$N_{f}$, where a is a constant and b measures the strength of the
correlation\cite{capela1,capela2}:
\begin{equation}
$b$ = \frac{\langle N_{f}N_{b}\rangle - \langle N_{f}\rangle \langle N_{b}\rangle}{\langle N_{f}^{2}\rangle - \langle N_{f} \rangle ^{2}}= \frac{D_{bf}^{2}}{D_{ff}^{2}}
\label{b}
\end{equation}
 $D_{bf}^{2}$ and $D_{ff}^{2}$ are the backward-forward and forward-forward dispersions respectively. The correlation strength given by Eq.(1) has the contributions from both short and long range sources. The long range part can be obtained by giving a large gap in rapidity between the forward and backward hemispheres.

\begin{figure}[hbp]
\centering
\includegraphics[width=0.8\textwidth]{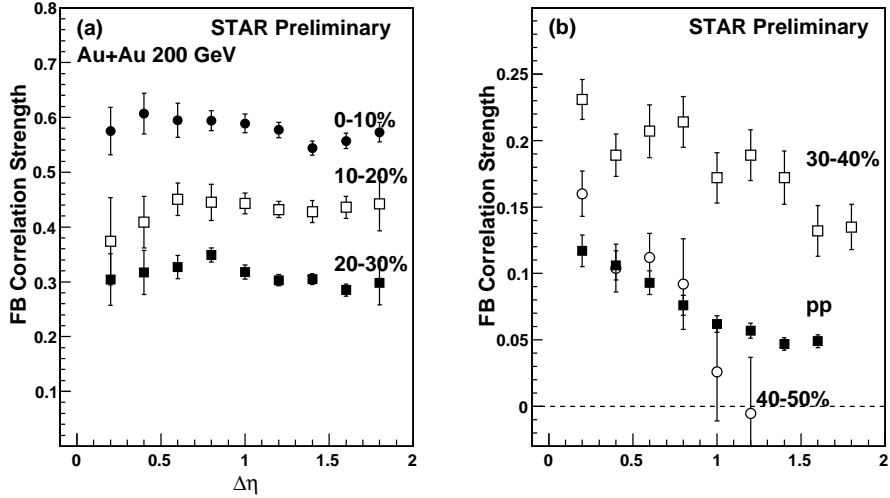}
\caption{ FB Correlation strength as a function of $\Delta\eta$ (a) for Au+Au at three centrality bins and (b) for \textit{p+p} and 30-40 and 40-50\% Au+Au. All errors are systematic.}
\label{FBAuandpp}
\end{figure}
 The STAR detector is most suited for forward-backward (FB) multiplicity 
correlations as it is symmetric about mid rapidity.
This is the first measurement of the FB correlation strength in nucleus-nucleus collisions at the highest RHIC energy.
The data utilized for this analysis is for Au+Au, Cu+Cu and {\it p+p} collisions at $\sqrt{s_{NN}}$ = 200 GeV at the Relativistic Heavy Ion Collider (RHIC), as measured by the STAR (Solenoidal Tracker at RHIC) experiment. 
The main tracking detector at STAR is the Time Projection Chamber (TPC) \cite{starnim}. All charged particles in the TPC pseudorapidity range -1.0$<\eta<$1.0 and  $p_{T} >0.15$ GeV/c were considered.  The collision events were part of the minimum bias dataset. The minimum bias collision centrality was determined by an off-line cut on the TPC charged particle multiplicity within the range -0.5$<\eta<$0.5. The forward-backward intervals were located symmetrically about midrapidity with the distance between bin centers ($\Delta\eta$) ranging from  0.2 to 1.8 with an interval of 0.2.

\begin{figure}[thbp]
\centering
%\vspace*{0.2cm}
\includegraphics[width=0.8\textwidth]{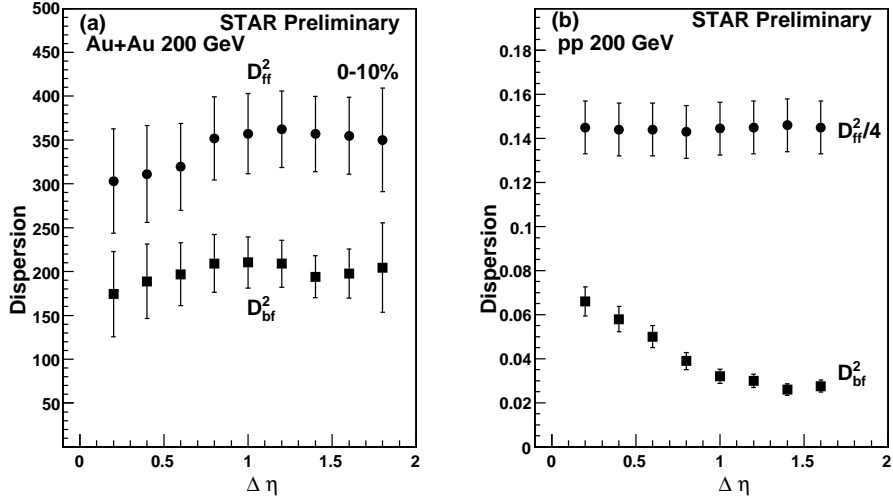}
%\vspace*{8pt}
\caption{ Backward-forward dispersion ($D_{bf}^{2}$) and  forward-forward dispersion ($D_{ff}^{2}$)  as a function of pseudorapidity gap $\Delta\eta$ (a) For Au+Au collisions at $\sqrt{s_{NN}}$ = 200 GeV and (b) for \textit{p+p} collisions. All errors are systematic.}
\label{DfAuandpp}
\end{figure}
Tracking efficiency and acceptance corrections were applied to each event. These were then used to calculate the backward-forward and forward-forward dispersions, $D_{bf}^{2}$ and $D_{ff}^{2}$, binned according to the STAR centrality definitions and normalized by the total number of events in
each bin.

The plot of FB correlation strength ($b$) as a function of the pseudorapidity gap is shown in Fig. \ref{FBAuandpp}(a) for the 0-10, 10-20 and 20-30\% central Au+Au events. 0-10\% being most central events. It is observed that the value of $b$ does
not change with the pseudorapidity gap. Fig. \ref{FBAuandpp}(b) shows $b$ as a function of $\Delta\eta$ for 30-40 and 40-50\%  Au+Au collisions along with the
{\it p+p} events. In this case $b$ decreases with the increasing $\Delta\eta$,
which is expected if there were only short range correlations. 
The centrality of the collision plays an important role in the growth of long range component of the total correlation strength. The magnitude of the LRC is quite large for the most central collisions when $ \Delta\eta >$ 1.0. Figure \ref{FBAuandpp}(b) shows that FB correlation strength in 40-50\% Au+Au falls faster with $ \Delta\eta $ as compared to {\it p+p} collisions. There is some hint of decreasing FB correlation strength with $\Delta\eta$ in 30-40\% Au+Au events as well.

 The plot of $D_{bf}^{2}$ and $D_{ff}^{2}$ as a function of the pseudorapidity gap is shown in Fig. \ref{DfAuandpp}(a) for the 0-10\% most central Au+Au events. It is observed that the value of $D_{bf}^{2}$ and $D_{ff}^{2}$ does not change with the pseudorapidity gap. Figure \ref{DfAuandpp}(b) shows $D_{bf}^{2}$, and $D_{ff}^{2}$, as a function of $\Delta\eta$ for the \textit{p+p} collisions. 
 Figures \ref{DfAuandpp}(a) and (b) show that change in  $D_{bf}^{2}$ with $\Delta\eta$ is quite different in 0-10\% Au+Au as compared to \textit{p+p}.  $D_{bf}^{2}$ falls with $\Delta\eta$ for \textit{p+p} and can be approximated by Gaussian or exponential function. This shows that FB correlation strength is controlled mainly by $D_{bf}^{2}$. 

\begin{figure}[thbp]
\centering
%\vspace*{0.5cm}
\includegraphics[width=0.8\textwidth]{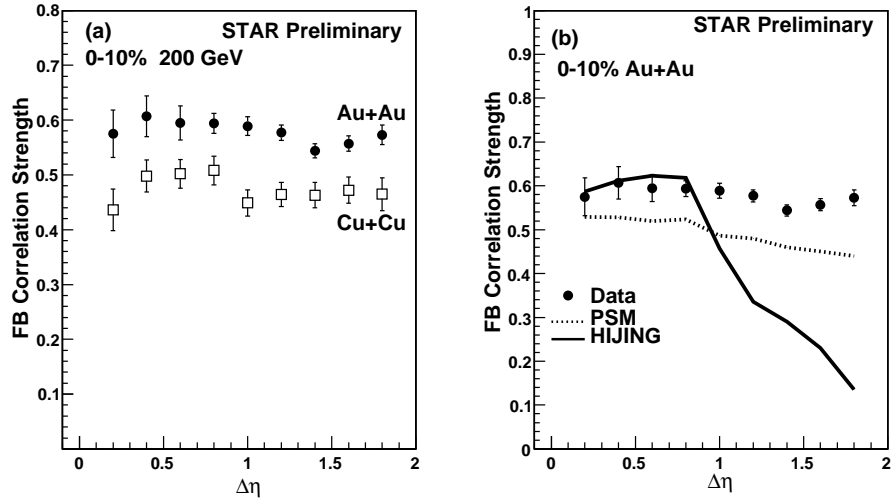}
%\centerline{\psfig{file=qm5_6_proc.eps,width=11cm}}
\vspace*{2pt}
\caption{ (a) FB corelation strength for 0-10\% Cu+Cu and Au+Au events. 
(b) Model comparison with data. The correlation strength is shown for HIJING  and the Parton String Model for the 0-10\% centrality in Au+Au collisions.}
\label{Models}
\label{FBCuandModel}
\end{figure}

 The FB correlation strength from the analysis of Cu+Cu at 
 $\sqrt{s_{NN}}$ = 200 GeV is shown in Fig. \ref{FBCuandModel}(a) along with the Au+Au for 0-10\% centrality. It is observed that the FB correlation strength is 
decreased by $ ~$15\% in going from Au+Au to Cu+Cu system. This suggests that system size does not have large effect on  FB correlation strength.

The 0-10\% results are also compared with phenomenological models HIJING \cite{hijing} and the Dual Parton Model (DPM) \cite{capela2}. Monte Carlo codes HIJING and the Parton String Model (PSM) \cite{capela2,amelin2,armesto3} were used to generate minimum bias events for Au+Au collisions at 200 GeV. The PSM is based on DPM \cite{capela2}. The variation of FB correlation strength  with $\Delta\eta$ is shown in Fig. \ref{FBCuandModel}(b) along with the experimental  value for 0-10\% central Au+Au collisions. HIJING predicts SRC with a large value of $b$ near midrapidity in agreement with the data. A sharp decrease is seen in FB correlation beyond the $\Delta\eta \sim$ 1.0. PSM has both short and long range correlations and is in qualitative agreement with the data. 

In the DPM, long range correlations are due to fluctuations in the number of elementary inelastic collisions and is given by the backward-forward dispersion \cite{capela2}:

\begin{equation}
\langle N_{f}N_{b}\rangle - \langle N_{f}\rangle \langle N_{b}\rangle \propto [\langle n^{2}\rangle - {\langle n \rangle}^{2}] {\langle N_{q-\overline{q}}\rangle}_{f}  {\langle N_{q-\overline{q}}\rangle}_{b}
\label{LRC}
\end{equation}
where the average multiplicities of $q-\overline{q}$ in the forward and backward regions is given by 
$ {\langle N_{q-\overline{q}}\rangle}_{f}$ and  
$ {\langle N_{q-\overline{q}}\rangle}_{b}$ 
respectively in each elementary inelastic collision. Eq. (\ref{LRC}) shows that the LRC is due to fluctuations in the number of elementary inelastic collisions. 
It is believed that the experimental observation of the LRC originates from these multiple partonic interactions. The CGC also  argues for the existence of a LRC in rapidity, similar to those predicted in DPM \cite{larry2}.

\end{document}